\documentclass[runningheads]{llncs}
\usepackage[colorlinks, citecolor=blue]{hyperref}
\usepackage{cite}
\usepackage{wrapfig}
\usepackage{amsmath,amssymb,amsfonts}
\usepackage{algorithmic}
\usepackage{graphicx}
\usepackage{textcomp}
\usepackage{xcolor}
\usepackage{xcolor}  % 颜色支持
\usepackage{lmodern}
\usepackage{soul}
\usepackage{marvosym}
\usepackage{amsmath,amssymb,amsfonts}
\usepackage{mathtools}
\usepackage{booktabs}
\usepackage{color}
\usepackage{diagbox}
\usepackage{textcomp}
\usepackage{array}
\usepackage{hyperref}
\usepackage{lipsum}
\usepackage{colortbl}
\usepackage{makecell}
\usepackage{booktabs,tabulary,multirow,overpic,xcolor}
\usepackage[ruled,vlined,linesnumbered]{algorithm2e}

\AtBeginDocument{\definecolor{tmlcncolor}{cmyk}{0.93,0.59,0.15,0.02}\definecolor{NavyBlue}{RGB}{0,86,125}}
\definecolor{color3}{rgb}{0.95,0.95,0.95}
\hypersetup{
    colorlinks=true,
    linkcolor=blue,  % 你可以修改这里设置其他类型的链接颜色
    urlcolor=violet,  % 这是超链接的颜色设置为粉色（violet 是一个接近粉色的颜色）
}

\usepackage[T1]{fontenc}
\usepackage{graphicx}
\begin{document}
\title{SecureNT: Smart Topology Obfuscation for Privacy-Aware Network Monitoring}
\author{Chengze Du\inst{1},
Jibin Shi\inst{2}, Hui Xu \inst{3,\textsuperscript{\Letter}}, Guangzhen Yao \inst{4, \textsuperscript{\Letter}}}
\institute{ 
\texorpdfstring{School of Cyberspace Security, Beijing University of Posts and Telecommunications, China  \newline} \email{\texttt{ducz0338@bupt.edu.cn}} \and
\texorpdfstring{School of Artificial Intelligence, Xidian University, China \newline}
\email{\texttt{zround@stu.xidian.edu.cn}} \and
\texorpdfstring{Smart Welfare Collaborative Research Center, Changchun Humanities and Sciences College, China \newline}
\email{\texttt{xuh504@nenu.edu.cn}} \and
\texorpdfstring{College of Science, National University of Defense Technology, China \newline}
\email{\texttt{yaoguangzhen@nenu.edu.cn}}
}
% \institute{ 
% \texorpdfstring{School of Cyberspace Security, Beijing University of Posts and Telecommunications, China  \newline}  \and
% \texorpdfstring{School of Artificial Intelligence, Xidian University, China \newline}
%  \and
% \texorpdfstring{Smart Welfare Collaborative Research Center, Changchun Humanities and Sciences College, China \newline}
%  \and
% \texorpdfstring{College of Science, National University of Defense Technology, China \newline}
% \email{\texttt{ducz0338@bupt.edu.cn, zround@stu.xidian.edu.cn, xuh504@nenu.edu.cn,  yaoguangzhen@nenu.edu.cn}}
% }

% \author{Anonymous authors}
% \institute{Anonymous institutes}
%
\maketitle              % typeset the header of the contribution
\begin{abstract}
Network tomography plays a crucial role in network monitoring and management, where network topology serves as the fundamental basis for various tomography tasks including traffic matrix estimation and link performance inference. The topology information, however, can be inferred through end-to-end measurements using various inference algorithms, posing significant security risks to network infrastructure. While existing protection methods attempt to secure topology information by modifying end-to-end measurements, they often require complex computation and sophisticated modification strategies, making real-time protection challenging. Moreover, these modifications typically render the measurements unusable for network monitoring, even by trusted users. This paper presents a novel privacy-preserving framework  that addresses these limitations. Our approach provides efficient topology protection while maintaining the utility of measurements for authorized network monitoring. Through extensive evaluation on both simulated and real-world networks, we demonstrate that our framework achieves superior privacy protection compared to existing methods while enabling trusted users to effectively monitor network performance. Our solution offers a practical approach for organizations to protect sensitive topology information without sacrificing their network monitoring capabilities. Source code is available at \href{https://gitee.com/Monickar/secure-nt}{https://gitee.com/Monickar/secure-nt}.

\end{abstract}
\section{Introduction}
Network tomography has emerged as a crucial technique for understanding and monitoring large-scale networks through end-to-end measurements\cite{he2021network-tomography, du2024identificationpathcongestionstatus}. By analyzing these measurements, network operators can infer internal network characteristics without requiring direct access to network elements. This non-intrusive approach has become increasingly important as networks grow in complexity and scale, particularly in scenarios where direct measurement of network components is impractical or impossible. Central to the effectiveness of network tomography is its reliance on accurate network structural information.

% Network topology, which describes the arrangement and connections between network elements, serves as the fundamental infrastructure that determines how network traffic flows and how performance metrics propagate through the network. Understanding and maintaining accurate topology information is essential for network operators to effectively manage their networks, troubleshoot problems, and optimize performance. The role of topology in network tomography extends far beyond simple connectivity mapping. Network topology serves as the essential foundation for various network tomography tasks, primarily in two critical areas: Origin-Destination (OD) flow estimation and link performance inference\cite{kakkavas2020review}. In OD flow estimation, topology information enables the construction of routing matrices that map observed traffic patterns to underlying network flows. For link performance inference, topology knowledge allows the decomposition of end-to-end measurements into individual link characteristics, making it possible to identify and locate performance bottlenecks or failures.

% Network topology, which describes the arrangement and connections between network elements, serves as the fundamental infrastructure that determines how network traffic flows and how performance metrics propagate through the network. 
Recent years have seen significant advancement in topology inference methods. Researchers have developed various classic approaches to infer network topology from end-to-end measurements\cite{EM2001, mle2002, ni2009efficient}. These inference techniques have become increasingly sophisticated, combining multiple measurement types and leveraging machine learning approaches\cite{deepnt, NeuralNT} to improve accuracy. 
\begin{figure}[t]
	\centering
	\includegraphics[width=\linewidth]{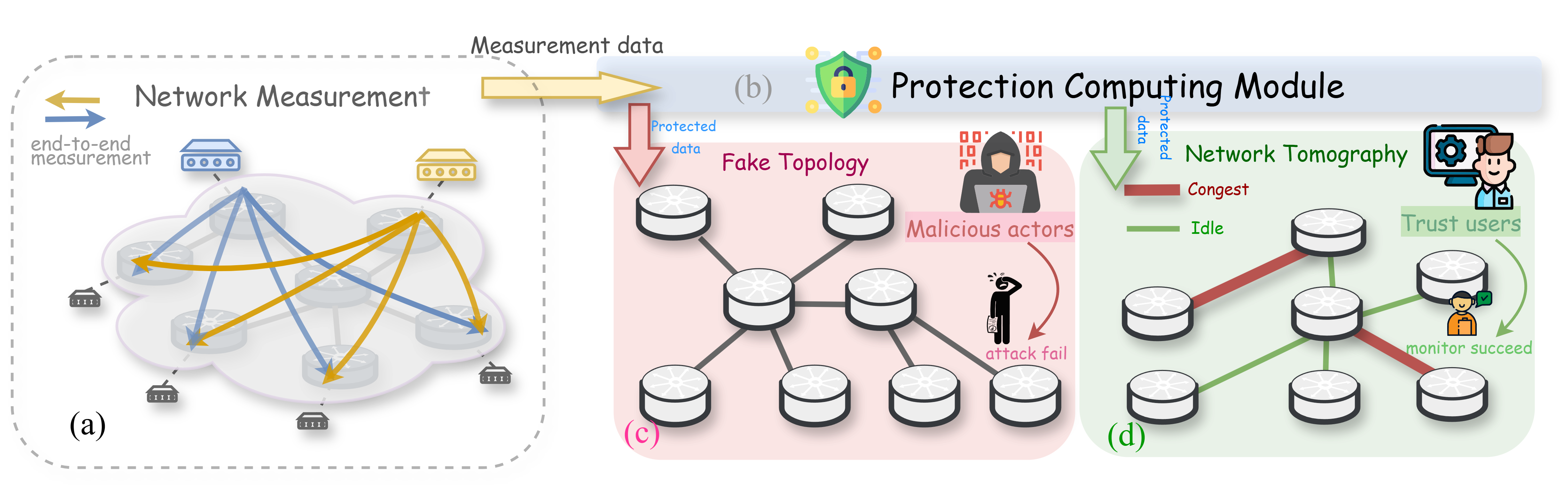}
	\caption{Overview of our framework: (a) Network Measurement component collecting end-to-end measurements across network paths. (b) Protection Computing Module that processes measurement data. (c) Fake Topology presented to malicious actors, causing their attacks to fail. (d) Network Tomography allowing trusted users to successfully monitor network status, identifying congested and idle links.}
	\label{Overview}
\end{figure}

However, the exposure of topology information presents significant security vulnerabilities that malicious actors could exploit. Malicious actors who can infer network topology potentially identify critical network components\cite{attack-1}, discover potential attack vectors\cite{attack-2}, or plan targeted disruptions of network services\cite{attack-3}. This vulnerability is particularly concerning as topology information can reveal the hierarchical structure of networks, including critical paths and potential single points of failure\cite{attack-4}. Furthermore, knowledge of network topology can facilitate various attacks, such as traffic analysis, denial of service attacks, or targeted infrastructure compromise\cite{attack-5}.

To counter these inference capabilities, several protection approaches have been proposed\cite{antitomo, Proto, Attack}. Current protection methods typically focus on modifying end-to-end measurements to obscure topology information. However, these approaches face significant limitations. Most notably, they often require complex computation to determine appropriate modifications, making real-time protection challenging. Additionally, these methods ignore the topology protection lies in the inherent trade-off between privacy and utility, which often significantly degrade the quality of network measurements in their attempt to protect topology information. This degradation poses a particular problem for network operators who need accurate measurements for legitimate monitoring and management tasks. The challenge becomes even more acute when considering environments where both trusted and untrusted users need to work with the same measurement data, but with enough accurate to classify whether the probe is malicuous.

Given these challenges, we propose a privacy-preserving framework and mechanism for network tomography (\textbf{SecureNT}, seen in Fig \ref{Overview}). First, the protection mechanism operates in real-time, providing immediate safeguarding of topology information as measurements are taken. This feature is critical for preventing temporal analysis attacks that could exploit delays in implementing protection. Second, our solution maintains measurement utility for trusted users, ensuring that operators and legitimate monitoring systems can accurately assess network performance and identify issues, even while topology information remains inaccessible to unauthorized users. Achieving this balance requires careful trade-offs between data modification and data usability. Third, the protection mechanism effectively prevents topology inference by attackers while remaining resilient to various inference techniques. This includes protection against both current methods and potential future approaches that might leverage advanced analysis techniques or combine different types of measurements. 

\textbf{\textit{Contribution}} The main contributions of this paper are as follows:
\begin{itemize}
    \item \textbf{\textit{Framework}} We propose a novel privacy-preserving framework for network tomography that effectively protects topology information while maintaining measurement utility.
    \item \textbf{\textit{Mechanism}} We design an efficient mechanism that provides real-time protection without requiring complex computation or sophisticated modification strategies.
    \item \textbf{\textit{Evaluation}} Through extensive evaluations, we demonstrate that our framework achieves superior protection while maintaining measurement utility for trusted users.
\end{itemize}

\section{Related Work}
Network topology inference and obfuscation represent two interrelated areas of network research. % While inference methods have evolved from traditional statistical approaches to sophisticated machine learning solutions, obfuscation techniques have emerged as crucial countermeasures against topology inference attacks. 

\textbf{\textit{Topology Inference}} Network topology inference has been extensively studied in network tomography, with early methods relying on additive metrics to formulate the problem as a linear inverse task, leveraging the known link-path relationships. Techniques such as Maximum Likelihood Estimation (MLE)\cite{mle2002} and Expectation Maximization (EM)\cite{EM2001} were employed for robust topology reconstruction, while algebraic methods like Systems of Linear Equations (SLE) effectively addressed scenarios with sparse measurements\cite{ni2009efficient}. Machine learning-based methods have further advanced the field by overcoming limitations of traditional approaches. For example, NeuTomography\cite{NeuralNT} and DeepNT\cite{deepnt} employ deep neural architectures to infer network structures and predict path performance without prior topology knowledge, offering enhanced scalability and adaptability to diverse network conditions. These advancements demonstrate the evolution from statistical and algebraic methods to more flexible, data-driven approaches for network topology inference.

\textbf{\textit{Topology Obfuscation}} Recent advancements in network topology obfuscation have emerged to counter various adversarial inference attacks using diverse strategies. Techniques such as AntiTomo\cite{antitomo} and Proto\cite{Proto} strategically manipulate end-to-end delay measurements to mislead attackers while preserving the real topology, whereas EigenObfu\cite{Eigenobfu} explicitly modifies graph structures to generate convincing fake topologies. While these methods primarily address static network representations, HBB-TSP\cite{HBB-TSP} addresses the high-order adaptability limitation by leveraging dynamic hypergraphs for real-time obfuscation of critical links. Complementary approaches such as NetHide\cite{Nethide} provide optimization-driven and SDN-based moving target defenses, respectively, while methods like PINOT\cite{Programmable} and Attack Graph Obfuscation\cite{Attack} focus on achieving an optimal balance between anonymity, routing efficiency, and deceptive resource allocation.

\section{Preliminaries and Problem Definition}
\subsection{Network Model and Performance Inference}
We consider a network represented as an undirected graph $\mathcal{G}=(\mathcal{N},\mathcal{L})$, where $\mathcal{N}$ represents the set of nodes (such as routers, switches, and hosts) and $\mathcal{L}$ denotes the set of links connecting these nodes. Within this network, source nodes generate data flows that traverse through the network to reach destination nodes. Each direct connection between two nodes is represented by a link $l \in \mathcal{L}$.

A path $p$ is defined as an ordered set of links that connect a source-destination pair, and we denote the set of all paths in the network as $\mathcal{P}$. The relationship between paths and links is captured by the routing matrix $\mathbf{R} \in \{0,1\}^{|\mathcal{P}| \times |\mathcal{L}|}$. If the element in the $i$-th row and $j$-th column of $\mathbf{R}$ is 1, it indicates that the $i$-th path contains the $j$-th link; otherwise, it is 0.

In network tomography, the goal is to infer link-level performance metrics $\mathbf{X} \in \mathbb{R}^{|\mathcal{L}|}$ from end-to-end path measurements $\mathbf{Y} \in \mathbb{R}^{|\mathcal{P}|}$. This relationship can be expressed as:
\begin{equation}
    y_{i}=\bigodot_{j=1}^{|\mathcal{L}|} R_{i j} \cdot x_{j} \Longrightarrow \mathbf{Y}=\mathbf{R} \bigodot \mathbf{X}
\end{equation}

where $\bigodot$ represents an operation that varies depending on the performance metric being considered. For additive metrics such as delay, $\bigodot$ represents summation, whereas for metrics like capacity, it could represent a minimum operation. Each element $y_i$ represents the end-to-end measurement on path $i$, and each element $x_j$ represents the performance metric (such as delay or capacity) on link $j$. This inference process critically depends on knowledge of the routing matrix $\mathbf{R}$, which is derived directly from the network topology. This fundamental relationship between topology and performance inference underlies both the utility of network tomography for legitimate monitoring and its vulnerability to topology inference attacks.

\subsection{Attacker Model and Topology Inference}
In our threat model, the attacker aims to infer the network topology $\mathcal{G}$ by analyzing end-to-end measurements $\mathbf{Y}$. Although the attacker lacks direct knowledge of the routing matrix $\mathbf{R}$ or the topology $\mathcal{G}$, they employ an inference algorithm $f(\cdot)$ to estimate the routing matrix $\hat{\mathbf{R}}$ and reconstruct the topology $\hat{\mathcal{G}}$. This inference process can be expressed as:
\begin{equation}
\hat{\mathcal{G}} = f(\mathbf{Y}), \quad \text{where } \hat{\mathcal{G}} = \left\{ \hat{\mathcal{N}}, \hat{\mathcal{L}}, \hat{\mathbf{R}} \right\},
\end{equation}

with $\hat{\mathcal{N}}$, $\hat{\mathcal{L}}$, and $\hat{\mathbf{R}}$ representing the inferred nodes, links, and routing matrix, respectively. The inference process leverages temporal correlations in measurements, statistical properties of end-to-end delays, and patterns in performance metrics across different paths to improve the accuracy of the reconstruction.  

The effectiveness of the attack is measured by comparing the inferred topology $\hat{\mathcal{G}}$ with the actual topology $\mathcal{G}$. This comparison uses a similarity measure defined as:
\begin{equation}
\text{Similarity}(\mathcal{G}, \hat{\mathcal{G}}) = s(\mathcal{G}, \hat{\mathcal{G}}) = 1 - \frac{G_0}{G_1 + G_2},
\end{equation}

where $s(\mathcal{G}, \hat{\mathcal{G}})$ is based on a graph edit distance-inspired metric.$G_0$ quantifies the cost of transforming the actual topology $\mathcal{G}$ into the inferred topology $\hat{\mathcal{G}}$, $G_1$ represents the cost of removing all nodes and links from $\mathcal{G}$ to obtain an empty graph, and $G_2$ denotes the cost of constructing $\hat{\mathcal{G}}$ from an empty graph. The similarity score  ranges from 0 to 1. A value of 1 indicates that the inferred topology $\hat{\mathcal{G}}$ is identical to the actual topology $\mathcal{G}$, whereas smaller values signify greater structural differences between the two graphs. This similarity measure offers a flexible and intuitive way to assess the success of topology inference attacks, where a high similarity score reflects the attacker's ability to accurately reconstruct the network, underscoring the importance of robust protection mechanisms.

\section{METHODOLOGY}
\subsection{Path Relationship-based Noise Injection}

To protect the real topology from inference attacks, we introduce a fake topology represented by another routing matrix $\mathbf{R'} \in \mathbb{R}^{|P| \times |L'|}$, where $L'$ denotes the fake links. A plausible fake link-delay vector $\mathbf{X'} \in \mathbb{R}^{|L'|}$ is generated based on properties like link lengths, and the fake path measurements $\mathbf{Y'} \in \mathbb{R}^{|P|}$ are computed as follows:
\begin{equation}
    \begin{aligned}
        x'_j = \frac{c}{l_j' + 1} \Rightarrow y_{i}'=\bigodot_{j=1}^{|\mathcal{L}'|} \mathbf{R'}_{ij} \cdot x'_{j} 
        \Rightarrow \delta_y = \mathbf{R'} \bigodot \mathbf{X'}
    \end{aligned}
\end{equation}

where $L_j' > 0$ is the length of the $j$-th fake link, and $c$ is a scaling constant. This process assigns smaller delays to longer links and larger delays to shorter links, mimicking realistic variability in network delays.

To ensure that the noise $\delta_y$ blends smoothly with the true path measurements $\mathbf{Y}$ and reduces statistical differences, it is scaled and smoothed using a mechanism $M $. The final modified measurements are expressed as:
\begin{equation}
    \tilde{\mathbf{Y}} = \mathbf{Y} + \delta_y^{\text{adj}} =  \mathbf{R} \bigodot \mathbf{X} + \alpha M \delta_y
\end{equation}

where $\alpha$ adjusts the noise magnitude, and $M$ serves as a proctection computing moduel, which two distributions A and B as inputs and iteratively applies gradient descent with a projection step to minimize the L2 loss between them. This is achieved using two input distributions: the fake measurements $\mathbf{R'} \mathbf{X'}$ and the reference measurements $\mathbf{R} \mathbf{I}$, where $\mathbf{I} \in \mathbb{R}^{|L'|}$ is a vector of ones. The adjusted noise $\delta_y^{\text{adj}}$ is computed as:
\begin{equation}
    \delta_y^{\text{adj}} = M ( \mathbf{R'} \bigodot \mathbf{X'},  \mathbf{R} \bigodot \mathbf{I})
\end{equation}

\begin{figure}[htb]
	\centering
	\includegraphics[width=0.9\linewidth]{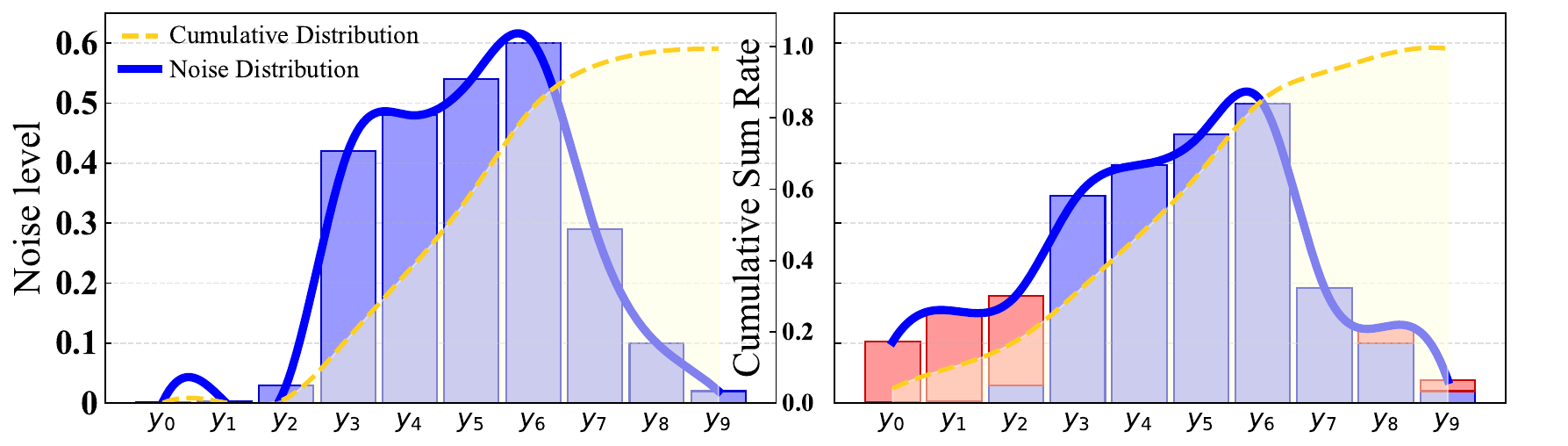}
	\caption{Noise between AntiTomo (left) and SecureNT (right). SecureNT implements path length-aware noise smoothing to achieve balanced noise distribution while preserving overall noise scale. Blue solid lines show the probability density of noise distribution, and yellow dashed lines represent cumulative distribution functions. The red shaded regions in the right subplot highlight the additional noise compensation introduced by SecureNT compared to AntiTomo.}
	\label{fig2}
\end{figure}

\begin{algorithm}[t]
    \SetKwInOut{Input}{Input}
    \SetKwInOut{Output}{Output}
    \caption{Protection Computing Moduel}
    \label{alg: distribution_adjustment}

    \Input{Initial distribution $\mathbf{Y}$, target distribution $\mathbf{RI}$, convergence threshold $\gamma$}
    \Output{Adjusted distribution $\mathbf{Y}'$}
    \BlankLine
    \tcc{\scriptsize{Initialize and Compute Initial Loss}}
    \BlankLine
    Set the initial total of $\mathbf{Y}$, $\mathbf{RI}$ to $\mathbb{S}(\mathbf{Y})$\, $\mathbb{S}(\mathbf{RI})$;
    
    Set $\alpha \leftarrow \frac{\mathbb{S}(\mathbf{RI})}{\mathbb{S}(\mathbf{Y})} \longrightarrow \mathbf{Y} \leftarrow \mathbf{Y} \times \alpha$ \tcc*{\scriptsize{Scale $\mathbf{Y}$ to match sum of $\mathbf{RI}$}}

    Set the initial loss $\mathcal{L}_{\text{initial}} \leftarrow \mathcal{L}(\mathbf{Y}, \mathbf{RI})$\;
    
    \BlankLine
    \tcc{\scriptsize{Iterative Gradient Descent with Projection}}
    \BlankLine
    \For{$t = 1$ \textbf{to} $T_{\text{max}}$}{
        % \tcc*{\scriptsize{Gradient descent step}}
        Compute the gradient: $\nabla \mathcal{L}(\mathbf{Y}, \mathbf{RI}) = 2 (\mathbf{Y} - \mathbf{RI})$\; 
        Update $\mathbf{Y}$: $\mathbf{Y} \leftarrow \mathbf{Y} - \eta \cdot \nabla \mathcal{L}(\mathbf{Y}, \mathbf{RI})$ 
        
        \BlankLine
        \tcc{\scriptsize{Projection to maintain the sum of Y}}
        \BlankLine
        Compute the current total: $\mathbb{S}(\mathbf{Y}') \leftarrow \mathbb{S}(\mathbf{Y})$\;
        Adjust $\mathbf{Y}$ : 
        $\mathbf{Y} \leftarrow \mathbf{Y} \times \frac{\mathbb{S}(\mathbf{RI})}{\mathbb{S}(\mathbf{Y}')}$\;
        
        \BlankLine
        \tcc{\scriptsize{Convergence Check}}
        \BlankLine
        Compute the current loss: $\mathcal{L}_{\text{current}} \leftarrow \mathcal{L}(\mathbf{Y}, \mathbf{RI})$\;
        \If{$\mathcal{L}_{\text{current}} \leq \gamma \cdot \mathcal{L}_{\text{initial}}$}{
            \textbf{break} \tcc*{\scriptsize{Converged, exit loop}}
        }
    }
    
    \Return{$\mathbf{Y}$}
\end{algorithm}

\subsection{Protection Objectives}
The noise injection mechanism is designed to achieve three intertwined objectives, which can be expressed as:
\begin{equation}
\mathbf{Y}_{best} = \min_{\mathbf{Y}'} \|\mathbf{Y}' - \mathbf{Y}\| - \lambda_1 d(\mathcal{G}, \mathcal{G}') + \lambda_2 \|\hat{\mathbf{X}}_t - \mathbf{X}\|
\end{equation}

where $\lambda_1$ and $\lambda_2$ are weighting factors that balance the importance of topology protection ($d(\mathcal{G}, \mathcal{G}')$) and performance inference accuracy ($\|\hat{\mathbf{X}}_t - \mathbf{X}\|$) against the need to preserve measurement fidelity ($\|\mathbf{Y}' - \mathbf{Y}\|$). First, measurement fidelity is preserved by minimizing the difference between the modified and original measurements, ensuring $\|\mathbf{Y}' - \mathbf{Y}\|$ remains small. Second, topology protection is maximized by increasing the difference between the real topology $\mathcal{G}$ and the topology $\mathcal{G}'$ inferred from the modified measurements, represented as $\max d(\mathcal{G}, \mathcal{G}')$, where $d(\cdot, \cdot)$ is a topology distance metric. Third, performance inference accuracy for trusted users is maintained by minimizing the error between the true link performance $\mathbf{X}$ and the inferred link performance $\hat{\mathbf{X}}_t$, ensuring $\|\hat{\mathbf{X}}_t - \mathbf{X}\|$ remains small. These objectives are carefully balanced through the design of the noise function $\eta_i(r_i)$, which scales with relationship strength to maximize topology protection while maintaining bounded noise to preserve fidelity and the utility of the measurements for trusted users.

\section{Evaluation}

\subsection{Datasets and Comparative Methods}
Our evaluation uses real-world network topologies from the Internet dataset Topology Zoo\cite{TopologyZoo}, which provides a diverse set of network configurations. We select 4 representative topologies, to ensure a thorough assessment of the proposed method. These topologies are summarized in Table \ref{TopologyMetrics}, covering various scales and structural complexities. 

\begin{table}[ht]
    \centering
    \caption{Characteristics of four real networks used for numerical simulations.}
    \scalebox{0.83}{
    \begin{tabular}{ccccc}
    \toprule
    \toprule
        \diagbox{Mec.}{Net.} & \textbf{CHINANET} & \textbf{AGIS} & \textbf{GANET} & \textbf{ERNET} \\ \midrule
        \textbf{\#Paths} & 17 & 14 & 15 & 12 \\ 
        \textbf{\#Links} & 21 & 18 & 17 & 13 \\ 
        \textbf{Avg.Hops} & 3.9 & 3.6 & 3.6 & 3.25 \\ 
        \textbf{Avg.Weights} & 4.3 & 2.8 & 3.1 & 3 \\ \bottomrule
    \end{tabular}
    }
    \label{TopologyMetrics}
\end{table}

We compare our approach with two state-of-the-art topology inference methods and two existing protection methods. For inference, we use \textbf{Maximum Penalized Likelihood (MPL)} as inference algorithm, which was a classical statistical approach that infers topology by maximizing the likelihood of observed end-to-end measurements. For protection, we include \textbf{AntiTomo}, which employs multi-objective optimization to manipulate delay measurements for topology protection, and \textbf{Proto}, a strategy-based method that introduces delays to create misleading topological impressions.

All methods were implemented in Python, with path measurements generated using NS-3, and experiments conducted on Linux.

\subsection{Topology Protection Effectiveness}

We futher evaluate topology protection effectiveness using Similarity Scores, which measure the similarity between the inferred and true network topologies. A lower similarity score indicates more effective protection, as it means the attacker's inferred topology differs more from the actual network structure.

\begin{figure}[ht]
	\centering
	\includegraphics[width=0.9\linewidth]{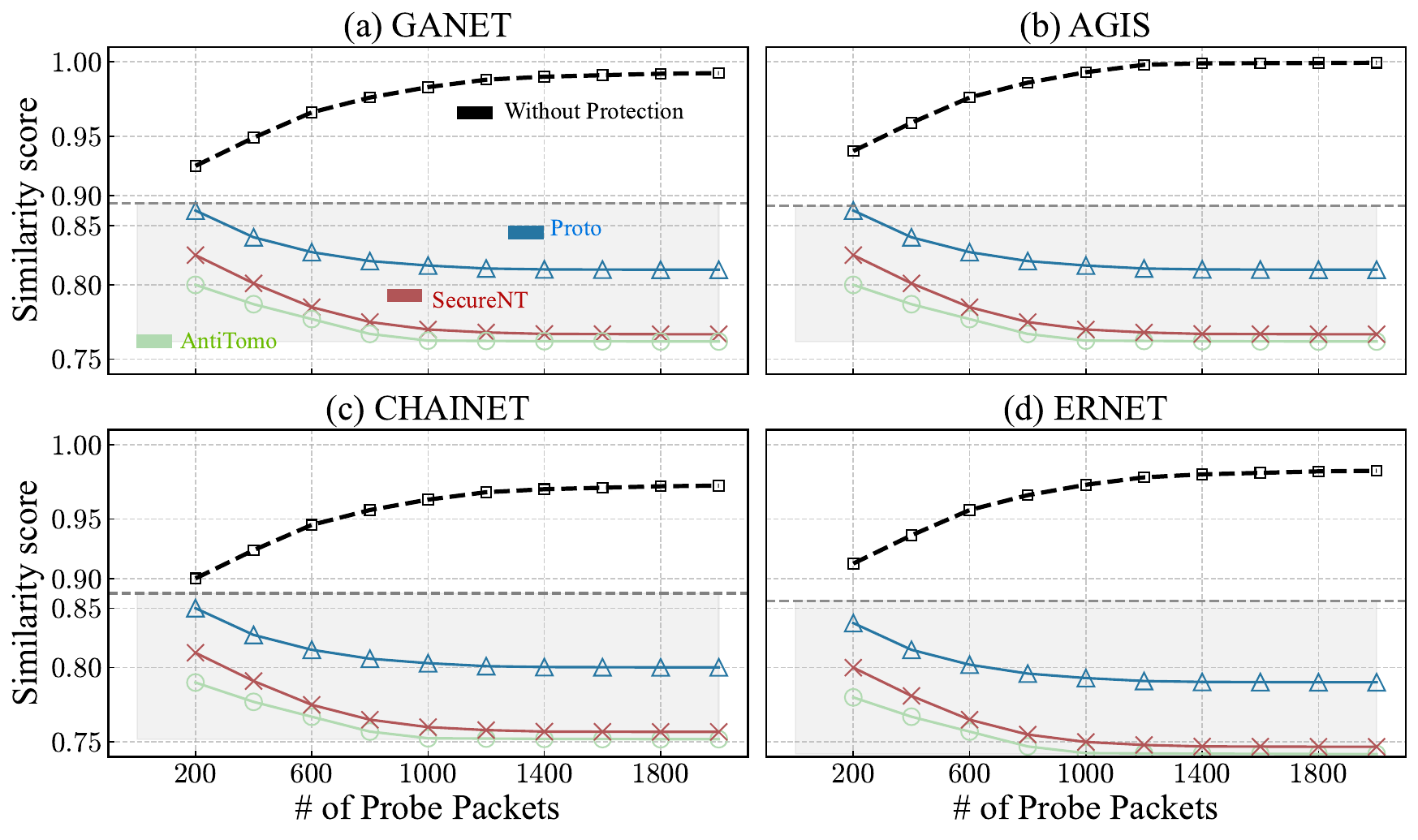}
        \caption{Comparison of topology protection effectiveness on four network topologies under varying numbers of probe packets. Lower similarity scores indicate better protection performance.}
        \label{fig3}
\end{figure}

Figure \ref{fig3} demonstrates the comparative performance of different protection methods across four network topologies (GANET, AGIS, CHAINET, and ERNET). The baseline case without protection (dashed black line) shows that attackers can achieve nearly perfect topology inference (similarity scores approaching 1.0) as the number of probe packets increases. Our proposed SecureNT method (red line) consistently outperforms Proto (blue line) and achieves performance comparable to AntiTomo (green line) across all evaluated networks.

For smaller networks like AGIS and GANET, SecureNT achieves an average topology similarity of 78.2\%, which is comparable to AntiTomo (77.8\%) and notably better than Proto (82.5\%). The protection effectiveness remains robust for larger networks like CHAINET and ERNET, where SecureNT maintains a similarity score of 77.7\%, matching AntiTomo's 77.4\% while outperforming Proto by 7.2\%. On average across all four topologies, SecureNT improves topology protection by 6.7\% compared to Proto, achieving an overall similarity score of 78.7\%.

Importantly, the graphs show that SecureNT's performance remains stable even as the number of probe packets increases from 200 to 1800, demonstrating consistent protection regardless of the intensity of probing attempts. This stability across different network sizes and probing intensities confirms that SecureNT provides reliable defense against topology inference attacks.

\subsection{Measurement Utility for Trusted Users}
\begin{figure}[htb]
	\centering
	\includegraphics[width=0.8\linewidth]{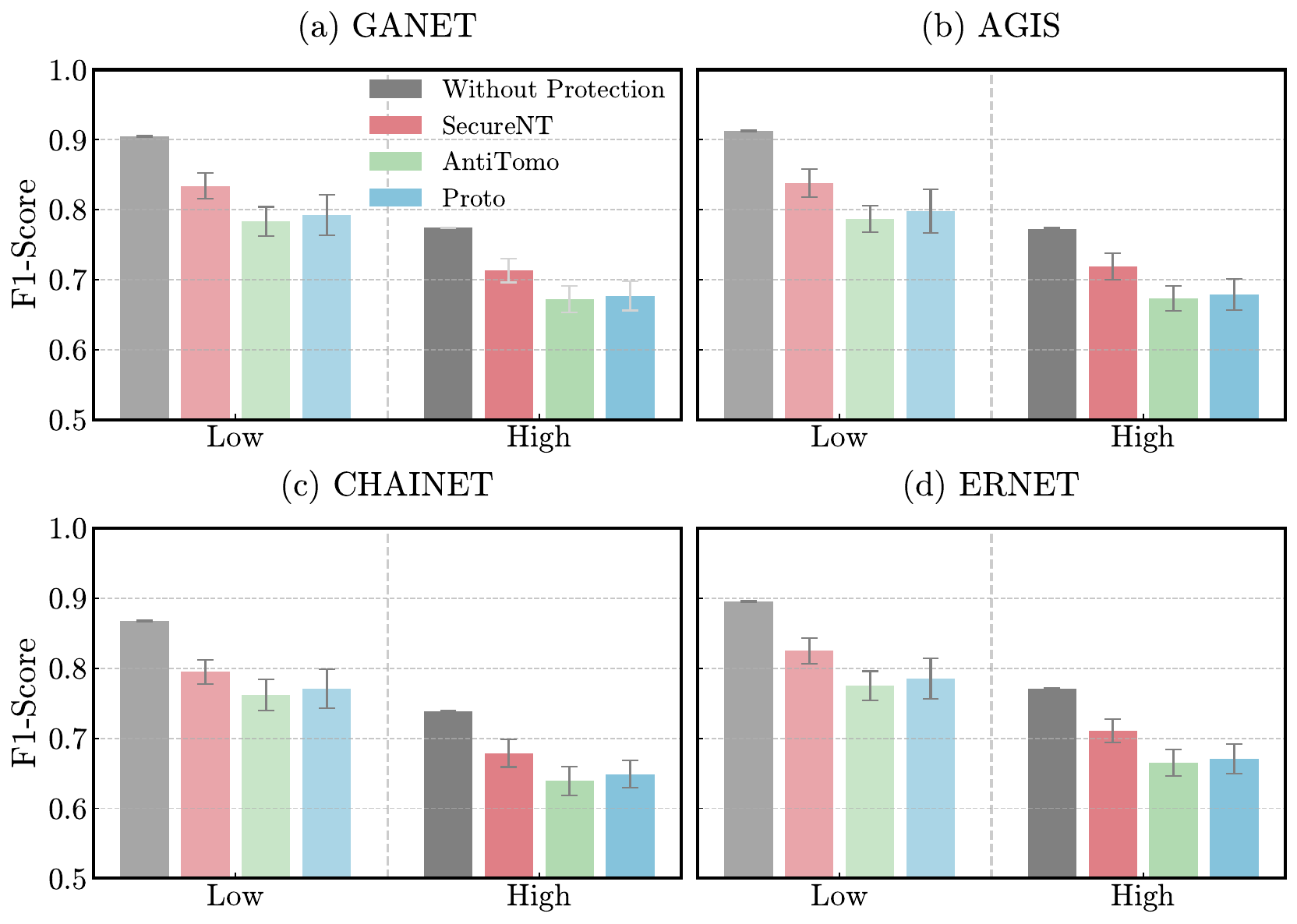}
        \caption{F1-scores of link congestion detection under low and high congestion levels across four network topologies, comparing Normal Tomography with three protection methods.}
        \label{fig4}
\end{figure}
To comprehensively evaluate how well our protection mechanism preserves measurement utility for trusted users, we analyze both binary congestion detection accuracy and continuous link performance inference capability.

For binary congestion detection, we employ the CLINK\cite{Nguyen_Thiran_2007_clink} algorithm, which determines link congestion status based on end-to-end measurements and network topology. Figure \ref{fig4} presents the F1-scores across four network topologies under both low and high congestion conditions. The baseline case (Without Protection) achieves F1-scores of approximately 0.87 in low congestion scenarios but drops to 0.75-0.77 under high congestion, reflecting the inherent challenges of congestion detection under heavy network load.

SecureNT demonstrates superior utility preservation compared to existing approaches. Under low congestion conditions, it maintains F1-scores of 0.83-0.85 for smaller networks (GANET and AGIS) and 0.80-0.82 for larger networks (CHAINET and ERNET), representing only a 7-10\% decrease from the baseline. In contrast, AntiTomo and Proto show larger performance degradation of 12-15\%. The advantage becomes more pronounced under high congestion scenarios, where SecureNT maintains F1-scores around 0.70-0.72 across all topologies, consistently outperforming other methods.

\begin{figure}[htb]
	\centering
	\includegraphics[width=0.85\linewidth]{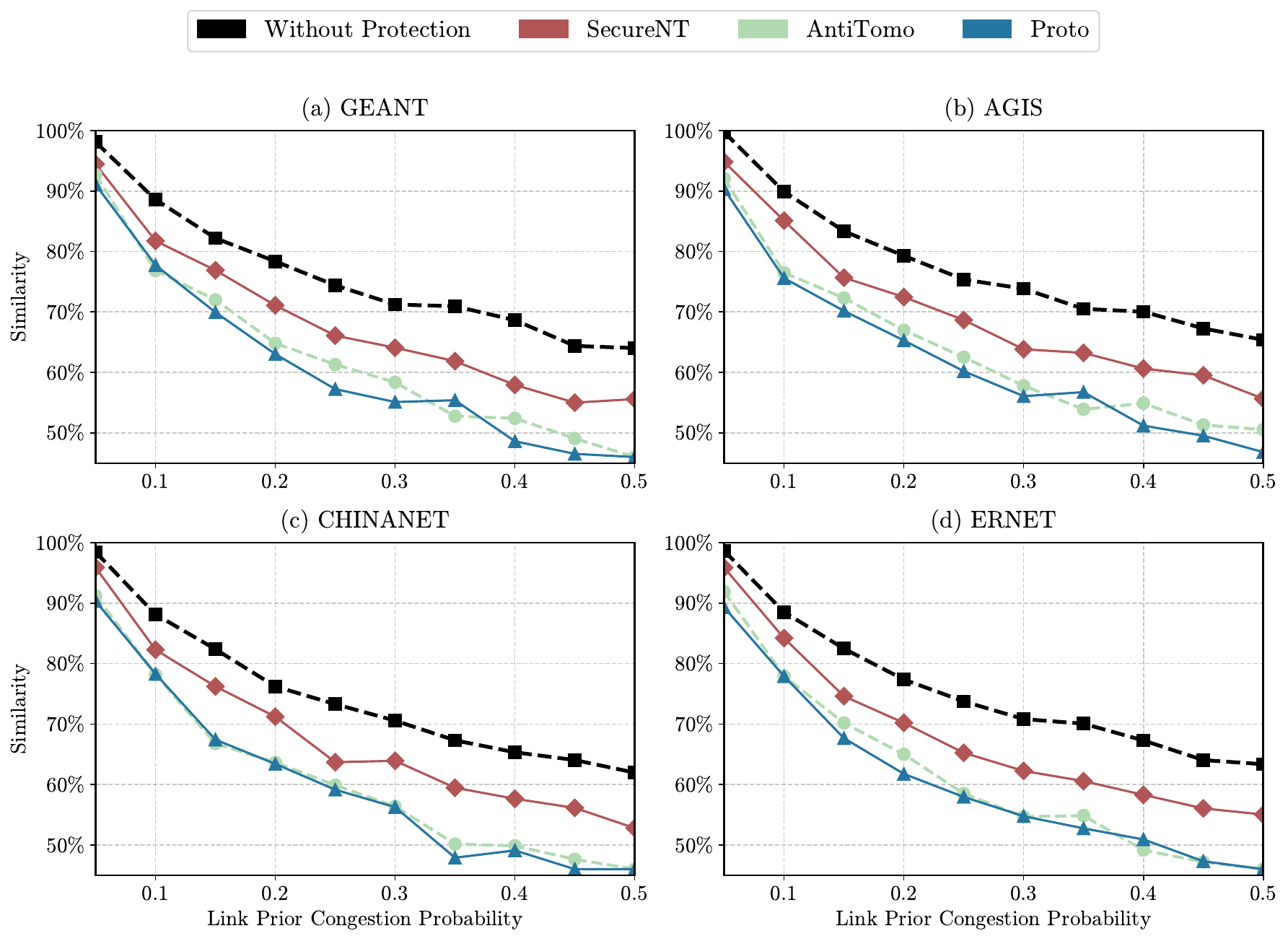}
        \caption{Impact of protection methods on link performance inference accuracy across four network topologies.}
        \label{fig: scores}
\end{figure}

For continuous link performance inference, we quantify accuracy using the normalized root mean square error (NRMSE):
$$\mathrm{NRMSE}=\sqrt{\frac{\sum_{m=0}^{M-1} |y[m]-\hat{y}[m]|^{2}}{\sum_{m=0}^{M-1} |y[m]|^{2}}}$$
where $y[m]$ represents the true link performance and $\hat{y}[m]$ represents the inferred performance. Figure \ref{fig: scores} shows how different protection methods affect RangeTomo\cite{range_nt} 's inference accuracy under varying link congestion probabilities. For moderate congestion probabilities (0.1-0.3), SecureNT maintains inference accuracy above 70\% across all networks, while AntiTomo and Proto drop below 65\%. At higher congestion probabilities (0.5), SecureNT preserves inference similarity around 55-60\% while other methods deteriorate to 45-50\%.

\section{Conclusion}

In this paper, we proposed a privacy-preserving framework for network tomography that achieves real-time topology protection while maintaining the utility of measurements for trusted users. Our approach effectively defends against both current and emerging topology inference techniques, ensuring robust privacy without compromising network monitoring capabilities. Extensive evaluations on simulated and real-world networks topology demonstrated the framework's superior privacy protection, usability,  making it practical for deployment in real-world scenarios. This work addresses critical challenges in network security and monitoring, providing a foundation for future advancements in balancing privacy and performance in large-scale networks.

\section*{Acknowledgement}
This work is supported by Changchun Humanities and Sciences College (2025KG03).

\vspace{-0.5cm}
\bibliographystyle{plain}  
\bibliography{references}

\vfill\pagebreak
\end{document}